# Elastocaloric signature of the excitonic instability in Ta$_2$NiSe$_5$

Elliott Rosenberg[1*], Joss Ayres-Sims[1], Andrew Millis[2], David Cobden[1], and Jiun-Haw Chu[1*]

[1]Department of Physics, University of Washington, Seattle, WA 98195, USA
[2]Department of Physics, Columbia University, New York, NY 10027, USA

*Correspondence to: erosenbe@uw.edu (E.R.) jhchu@uw.edu (J.-H.C).

On cooling through a temperature $T_S$ of around 324 K, Ta$_2$NiSe$_5$ undergoes a transition from a semimetallic state to one with a gapped electronic spectrum which is suspected to be an excitonic insulator. However, at this transition the structure also changes, from orthorhombic to monoclinic, leaving open the question of whether it is driven primarily by excitonic ordering or by a lattice instability. A lattice instability of this symmetry would correspond to softening of a B$_{2g}$ optical or acoustic phonon mode. Here, we report that elastocaloric measurements of Ta$_2$NiSe$_5$ with induced B$_{2g}$ strain reveal a thermodynamic susceptibility described by a Curie-Weiss law with a Curie temperature $T^*$ of 298 K. The fact that $T^*$ is close to $T_S$ rules out the possibility that the B$_{2g}$ acoustic mode is responsible for the transition. Since prior Raman measurements have shown minimal softening of the B$_{2g}$ optical mode as well, our finding strengthens the case that the transition is largely excitonic in nature. Our work underscores the potential of using strain as a tool for separating electronic and lattice contributions in phase transitions.

The excitonic insulator (EI) phase was first proposed in the 1960s as a possible ground state of a near zero-gap semiconductor or semimetal. In the EI, valence band holes and conduction band electrons hybridize to form excitons in chemical equilibrium which at sufficiently low temperature enter an ordered state, resulting in opening of an electronic gap[1–7]. Although many EI candidates have been examined, to date there has been no undisputed confirmation of an EI state in a bulk crystal. The main reason for this is that the excitonic order parameter always breaks a crystal symmetry and is thus is inevitably coupled to a lattice distortion[8]. As a result, the transition to the low temperature state is necessarily of mixed electronic and lattice character. Additionally, in most EI candidates, such as 1T-TiSe$_2$ and TmSe$_{0.45}$Te$_{0.55}$, the lattice distortion takes the form of a charge density wave (CDW) whose wavevector $q$ is the nesting vector between the conduction and valence band edges[9,10]. This makes it particularly difficult to separate electronic and lattice effects, to determine which dominates or how strong their coupling is.

In one particularly promising EI candidate, the situation is different[11,12]. Ta$_2$NiSe$_5$ exhibits a phase transition at $T_S \approx 324$ K[11] from a higher temperature semimetallic phase to a lower temperature semiconducting phase with a gap of ~0.16 eV as inferred from optical measurements[13]. At the same time, it undergoes a small lattice deformation changing its structure from orthorhombic to monoclinic[13]. As with other EI candidates, this raises the question of whether the transition is predominantly structural[14–16] or excitonic[12,17–20] in origin. Importantly, however, in Ta$_2$NiSe$_5$ the lattice deformation is not a translation-symmetry breaking CDW but rather a $q = 0$ mirror-symmetry breaking shear deformation. Since this shear deformation is conjugate to an experimentally controllable shear stress, we can *exploit* it to *probe* the transition by investigating the response to such stress. We do this by measuring the elastocaloric effect (ECE), which is proportional to the strain-induced entropy change of the sample and thus is sensitive to fluctuations of the order parameter on approaching a phase transition. From its temperature dependence we can derive vital information about the mechanism of the transition.

The structure of Ta$_2$NiSe$_5$ is shown in Fig. 1a. It contains linear Ta$_2$-Ni chains along the a-axis covalently bonded into layers that are van-der-Waals stacked along the b-axis. The structural change at the transition is essentially a 0.6° shear deformation of each layer (i.e., in the a-c plane) that produces a parallel relative shift of the chains. This change from orthorhombic to monoclinic breaks the mirror symmetry perpendicular to the c-axis and belongs to the $B_{2g}$ irreducible representation of the $D_{2h}$ point group of the orthorhombic structure. In the orthorhombic state the conduction and valence bands overlap along the high symmetry momentum direction because their hybridization is forbidden by the mirror symmetry[8]. In the monoclinic state the broken symmetry allows the bands to hybridize and a gap to open (Fig. 1b). In the pure excitonic insulator picture, the transition is driven by an excitonic mode, i.e., an instability towards formation of a macroscopic density of excitons. This results in band hybridization that would break the mirror symmetry even absent nuclear displacements[8,17]. However, there are also two phonon modes which could drive the transition: a $B_{2g}$ optical mode that involves the shearing of the tantalum cages around the nickel atoms within the chain[14,21], and a $B_{2g}$ acoustic mode which corresponds to the shear deformation of the entire lattice. These three possible modes of instability—excitonic, optical and acoustic—are shown schematically in Figure 1c.

Applying $B_{2g}$ stress induces $B_{2g}$ strain, $\varepsilon_{B2g}$ (measured relative to the orthorhombic structure), that is also the order parameter of the acoustic mode at $q \to 0$. Since all the relevant modes have the same $B_{2g}$ symmetry, they must couple to each other bilinearly. The optical and excitonic modes cannot be separated when inducing $\varepsilon_{B2g}$, so we lump them into a single combined "non-acoustic" order parameter $\phi$ and attempt to interpret the measurements using a simple Ginsburg-Landau free energy density of the form

$$F = \tfrac{1}{2}a_0(T - T^*)\phi^2 - \lambda\phi\varepsilon_{B2g} + \tfrac{1}{2}C\varepsilon_{B2g}^2 , \tag{1}$$

where $\lambda$ is a coupling coefficient and $a_0$ and $C$ are positive constants ($C$ is the bare shear modulus). This form of $F$ generates a phase transition driven by the non-acoustic modes which occurs at temperature $T^*$ when the coupling vanishes ($\lambda = 0$), and at a higher temperature, $T_S = T^* + \lambda^2/(a_o C)$, for finite coupling. We will show below that our measurement of the ECE determines the strain susceptibility of $\phi$ to have a Curie-Weiss temperature dependence, consistent with the assumptions of Eq. 1, that allows a determination of $T^*$ which is found to be only ~10% below $T_S$. This self-consistently validates the use of Eq. 1 and so proves that the non-acoustic modes do indeed drive the transition.

An analogous argument was the basis for the conclusion that the nematicity seen in the iron pnictide superconductors is primarily of electronic origin[22,23]. There, the strain susceptibility was measured via the strain-induced resistivity anisotropy (elastoresistivity) and found to exhibit a Curie-Weiss temperature dependence with a Curie temperature only slightly below the observed nematic transition temperature. While elastoresistivity is a powerful technique, it is not a thermodynamic measure of the strain susceptibility as it relies on the coupling of electrical transport coefficients with the order parameter. In contrast, the recently developed elastocaloric effect measures temperature changes and hence is a thermodynamic technique which directly probes the strain susceptibility of $q = 0$ symmetry breaking transitions[24–27].

The elastocaloric effect refers to the change of temperature of the sample induced by strain in the adiabatic limit. In the AC elastocaloric technique[24], a modulated strain with a DC offset component $\varepsilon_o$ and a small oscillating AC component $d\varepsilon$ is induced in the sample, and the resulting oscillating temperature change $dT$ is measured. The frequency is chosen to be higher than the thermal relaxation rate to the environment so that the measurement is effectively adiabatic, but low enough that the thermometer tracks the sample temperature. To induce the relevant $B_{2g}$ strain, $\varepsilon_{B2g}$ (see Methods for more details), the crystal is oriented so that the uniaxial stress is applied at 45° to the a-axis in the a-c plane. Experimentally, this is accomplished by gluing a crystal across a voltage-controlled gap in a commercial strain cell[28], as shown in Fig. 2a, and the magnitude of the induced uniaxial strain $\varepsilon_{xx}$ is estimated by measuring the gap displacement using a capacitor built into the strain cell. In this geometry, $\varepsilon_{B2g} = \alpha\varepsilon_{xx}$, where $\alpha$ is determined by the relevant Poisson ratio $\gamma'$ via $\alpha = \frac{1+\gamma'}{2}$ (See Supplement). A thermocouple attached to the middle of the crystal is used to detect the corresponding temperature change arising from the AC component of strain. The experimentally measured elastocaloric (EC) coefficient, $\eta = dT/d\varepsilon_{xx}$, can be shown in the adiabatic limit to be proportional to the strain derivative of the isothermal entropy $S$ of the sample,

$$\eta(\varepsilon_o, T) = \left(\frac{\partial T}{\partial \varepsilon_{xx}}\right)_S = -\frac{T}{C_\varepsilon}\left(\frac{\partial S}{\partial \varepsilon_{xx}}\right)_T, \qquad (2)$$

where the derivative of entropy and the constant-strain heat capacity $C_\varepsilon$ are evaluated at strain $\varepsilon_o$ and temperature $T$.

If the form of $F$ in Eq. 1 is valid, i.e., if the non-acoustic mode $\phi$ is the primary order parameter in the transition, then when $T > T_S$ both order parameters, $\phi$ and $\varepsilon_{B2g}$, vanish at zero stress. However, when shear strain $\varepsilon_{B2g}$ is externally induced, $\phi$ also becomes nonzero due to the bilinear coupling. This causes a decrease in the isothermal entropy quadratic in $\varepsilon_{B2g}$ whose magnitude depends on the strain susceptibility, defined as

$$\chi = \left(\frac{\partial \phi}{\partial \varepsilon_{B2g}}\right)_T.$$

Using Eq. 2 and $S = -\left(\frac{\partial F}{\partial T}\right)_\varepsilon$, where $F$ is described in Eq. 1, one finds that

$$\left(\frac{\partial S}{\partial \varepsilon_{B2g}}\right)_T = -\lambda \varepsilon_{B2g} \frac{d\chi}{dT}, \qquad (3)$$

with the susceptibility showing Curie-Weiss behavior,

$$\chi = \frac{\lambda/a_0}{T - T^*}. \qquad (4)$$

Combining Eqs. 2 and 3 gives the following expression for the EC coefficient at $\varepsilon_{xx} = \varepsilon_o$:

$$\eta(\varepsilon_o, T) = -\frac{\lambda a^2 T}{C_\varepsilon}\frac{d\chi}{dT}\varepsilon_o. \qquad (5)$$

It diverges at $T = T^*$ due to the divergence of $\chi$. If, on the other hand, Eq. 1 is invalid and the transition is instead driven by the acoustic mode, then no feature is expected in the EC coefficient near the transition.

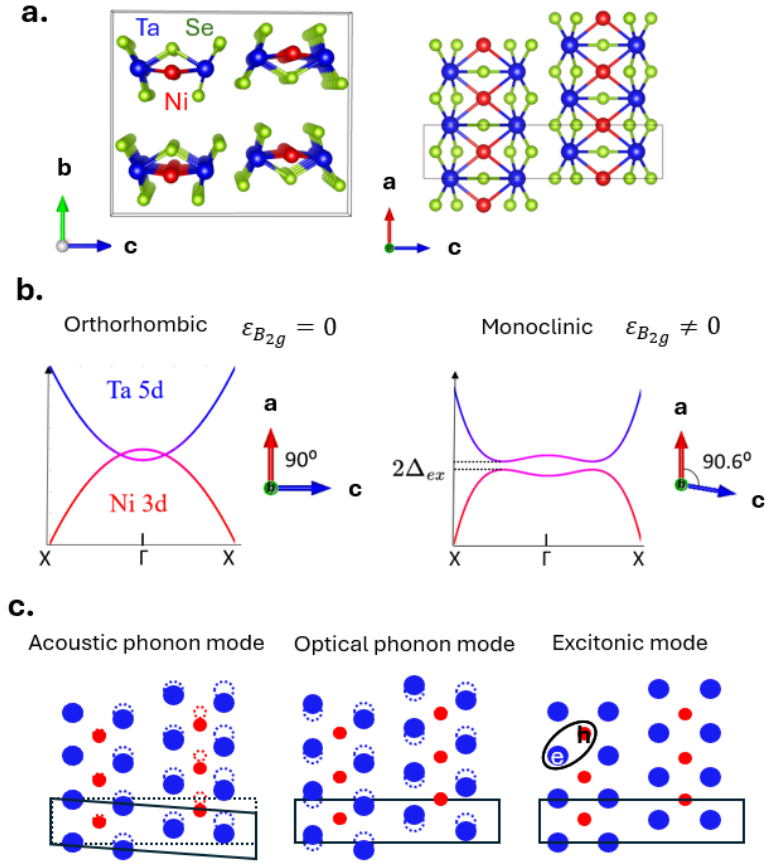

**FIG. 1. Structure, bands and relevant modes near the phase transition in Ta₂NiSe₅. a.** Depictions of the crystal structure looking along the chains (left) and at a layer from above (right). **b.** Schematics of the band structure in the high-temperature orthorhombic (semimetallic) and low-temperature monoclinic (gapped) phase, indicating the relevant crystal angles. **c.** Schematics of the three $B_{2g}$ modes that could drive the phase transition: the acoustic phonon mode; the optical phonon mode; and the excitonic mode. Inducing shear strain $\varepsilon_{B_{2g}}$ is equivalent to fixing the order parameter of the acoustic mode.

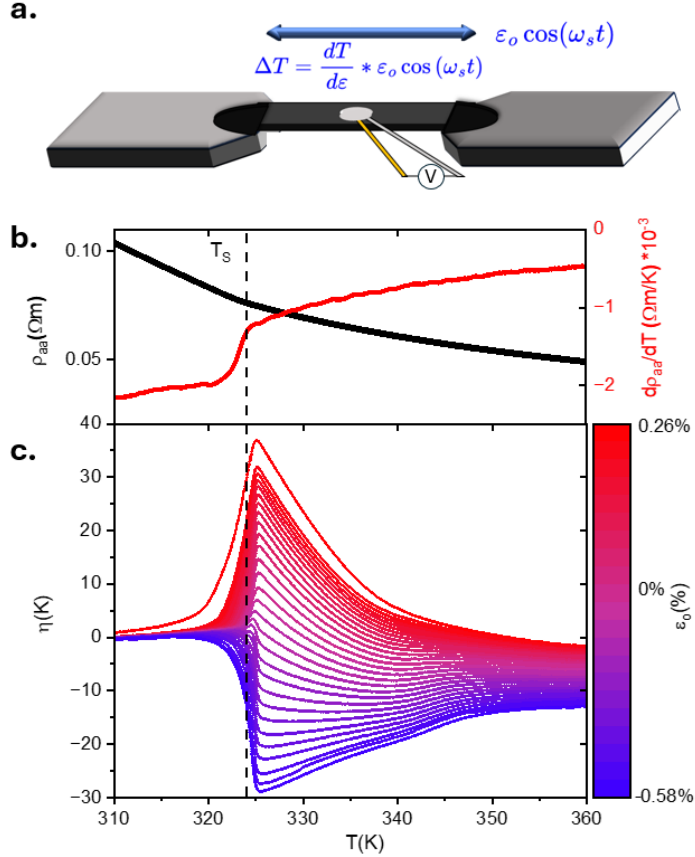

**FIG. 2. Elastocaloric (EC) coefficient measurements. a.** A Ta$_2$NiSe$_5$ crystal is glued across the titanium plates of a strain cell so as to induce strain at 45 degrees to the crystal a-axis. Modulating the strain changes the isothermal entropy of the sample, inducing a temperature change measured using a thermocouple (a junction between two different-metal wires) attached to the sample. The EC coefficient $\eta$ is the ratio of the magnitude of the temperature oscillations to that of the strain modulation ($\approx 0.005\%$). **b.** Resistivity vs temperature (black) and its derivative (red) in an unstrained sample. The step in the derivative occurs at the structural phase transition temperature, here $T_S = 324$ K. **c.** Measurements of the EC coefficient made while sweeping temperature at a series of offset strains $\varepsilon_o$, indicated by color.

Figure 2b shows the temperature dependence of the resistivity of our Ta$_2$NiSe$_5$ sample. The structural phase transition temperature $T_S$ at zero strain can be identified by a kink in the resistivity. Figure 2c shows measurements of the EC coefficient vs temperature at a series of values of the offset strain $\varepsilon_o$ ranging from approximately +0.25% tensile (red) to -0.56% compressive (blue). The zero of $\varepsilon_{xx}$ is taken to be the value of $\varepsilon_o$ where the EC response is at a minimum. Its behavior is immediately seen as qualitatively consistent with Eq. 5 (which hence justifies the relevance of the free energy in Eq. 1): it shows signs of diverging at some temperature below $T_S$, and it depends strongly on $\varepsilon_o$, changing sign as $\varepsilon_o$ goes from tensile to compressive.

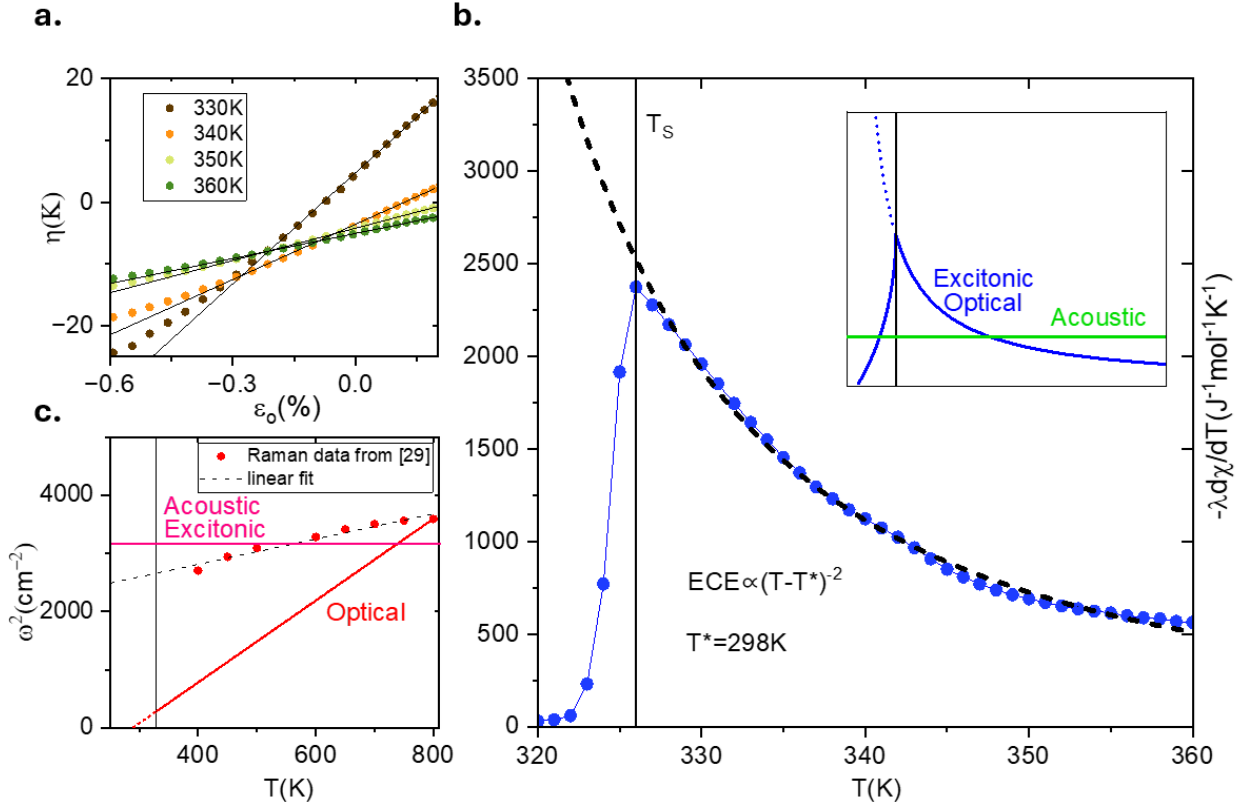

**FIG. 3. Determining the driving mechanism of the phase transition in Ta$_2$NiSe$_5$. a.** EC coefficient $\eta$ vs strain $\varepsilon_o$ at selected temperatures $T$ showing linear fits. **b.** Plot of $-\lambda \frac{d\chi}{dT}$ extracted from the slope $\left(\frac{\partial \eta}{\partial \varepsilon_0}\right)_T$ as a function of $T$ (see text.) The dashed curve is a fit of the form $\lambda^2/a_0 (T-T^*)^{-2}$ to the points at $T > T_S$. Inset is a schematic comparing the expected $T$ dependence of $-\lambda \frac{d\chi}{dT}$ for a phase transition driven by the acoustic phonon mode (green) with that for the excitonic/optical phonon modes (blue). **c.** The square of the frequency of the optical phonon mode as a function of temperature replotted from ref.[29]. Solid lines show the expected temperature dependence for a phase transition driven by optical phonon mode (red) and excitonic/acoustic phonon mode (pink).

To analyze the results quantitatively we note that, according to Eq. 5, at a fixed $T$ above $T_S$, $\eta$ is proportional to $\varepsilon_0$, its slope being $\left(\frac{\partial \eta}{\partial \varepsilon_0}\right)_T = -A\lambda \frac{d\chi}{dT}$ where $A = \alpha^2 T/C_\varepsilon$. Indeed, $\eta$ is found to be roughly linear in $\varepsilon_o$ at temperatures above $T_S$, as shown in Fig. 3a. The best-fit slope, divided by the factor A which was determined from heat capacity data in Ref. [13] and self-consistent calculations of α described in the Supplement, is plotted vs temperature in Fig. 3b. Since (from Eq. 4) $\chi = \frac{\lambda/a_0}{T-T^*}$, we fit the data for $T > T_S$ to the form $\lambda^2/a_0 (T-T^*)^{-2}$, treating $\lambda^2/a_0$ and $T^*$ as constant fitting parameters. This yields the black dashed curve, with $T^* = 298$ K with an estimated uncertainty of $\pm 1$ K. The high quality of the fit validates the assumptions behind Eq. (1). In addition, the closeness of $T^*$ to $T_S$, with $(T_S - T^*)/T_S \approx 0.1$, implies that the contribution to the phase transition from the coupling to the acoustic mode is small.

Having thus established that the transition is driven by some combination of the optical phonon and excitonic modes, lumped into a single order parameter $\phi$ whose fluctuations dominate the EC response, we now consider the evidence for the distinction between these two modes. If an instability of the optical mode drives the transition, as suggested by density functional theory calculations[14], the phonon frequency $\omega_o$ should soften substantially above the transition. For example, above the phase transition in BiVO$_4$ $\omega_o^2$ decreases linearly with $T$ and extrapolates to zero at 365 K which is ~70% of the transition temperature[30]. In Ta$_2$NiSe$_5$, however, there is very little softening[17,18,21]; Raman measurements from Ref. [29], replotted in Fig. 3c, show $\omega_o^2$ for the $B_{2g}$ optical phonon decreasing only slowly and extrapolating to zero at $T_o = -800$ K, very far below $T_S$. This rules out the optical mode as the lone driving factor and thereby implies that the excitonic mode is important in the transition in Ta$_2$NiSe$_5$.

However, the Raman linewidth is found to broaden considerably above $T_S$, indicating coupling of the optical mode to an electron-hole continuum[17,29]. In addition, recent time-resolved ARPES experiments[16] showed softening and recovery of the gap on the timescale of phonons, while other ARPES measurements revealed a large electron-phonon coupling[31]. A scenario consistent with these findings is that there is strong coupling between the optical phonon and excitonic modes[32], which can allow a phase transition even when neither mode softens (see Supplement), analogous to a co-operative Jahn-Teller effect[33].

In summary, our elastocaloric measurements, taken together with the literature, lead us to conclude that the instability driving the phase transition in the excitonic insulator candidate Ta$_2$NiSe$_5$ does indeed involve excitonic ordering, though likely strongly coupled to an optical phonon mode. They also show that shear strain can induce, and thus control, excitonic order above the phase transition temperature, producing a very clear characteristic thermodynamic response. From a more general perspective, the work demonstrates the value of strain-based techniques for determining the nature of phase transitions with a variety of $q = 0$ structural changes, including ones that break mirror symmetries.

**Methods**

For the elastocaloric measurements, a commercial Razorbill CS-100 strain cell was used to apply strain to the samples, which were cut 45 degrees from the a-axis to be approximately 1.5 mm x 0.4 mm x 0.02 mm in size. The samples were secured between two sets of mounting plates using Stycast 2850FT Epoxy, which were screwed into the strain cell, to have a gap of approximately 0.7 mm. An AC voltage of 2.5V RMS at 17 Hz was applied to the outer piezoelectric (PZT) stacks of the strain cell, corresponding to applying an AC displacement of the sample of approximately 0.005% of its length. This frequency was experimentally determined by measuring the elastocaloric signal at 330 K for frequencies in the range of 10-100 Hz and choosing the frequency with the largest response. This implied the frequency was at the plateau of the relevant thermal transfer function, which did not observably shift in the temperature range measured[24]. DC voltages were applied to the inner PZT to reach a strain range of 0.7%. To approximate the strain in the sample, a capacitor built into the strain cell was measured to determine the relative displacement of the sample plates, which was divided by the length of the gap. This is not fully accurate as it assumes 100% strain transmission. Strain transmission at high temperatures (350 K) decreases due to the softening of the epoxy used, but this was taken into account by measuring a strain gauge with a constant gauge factor glued in an identical way for use as a calibration.

The temperature fluctuations in the sample induced by the AC strain were measured using a home-made Type E (Chromel-Constantan) thermocouple. The chromel and constantan wires (50$\mu$m diameter) were thermally anchored to an outer part of the strain cell and silver pasted to the sample together. The voltage between the two wires was measured using an SRS860 lock-in amplifier at the frequency of the strain being induced to obtain the amplitude of the temperature fluctuations.

Transport measurements were performed simultaneously with elastocaloric measurements by sputtering gold pads on the sample and electrically connecting gold wires to them with silver paste in a four-probe measurement geometry before the sample was glued to the strain cell. Care was taken to not short the voltage pads with the thermocouple.

**Acknowledgments:** We thank G. Blumberg, R. M. Fernandes and Y. He for valuable discussion. **Funding:** This work was solely supported as part of Programmable Quantum Materials, an Energy Frontier Research Center funded by the U.S. Department of Energy (DOE), Office of Science, Basic Energy Sciences (BES), under award DE-SC0019443. **Author contributions:** E.R. and J.A-S. grew the samples E.R. and J.A-S. did the experiments. E.R. analyzed the data. A.M. provided theoretical support. J.-H.C. supervised the project. All authors contributed extensively to the interpretation of the data and the writing of the manuscript. **Competing interests:** Authors declare no competing interests. **Additional Information:** Correspondence and requests for materials should be addressed to E.R. and J.-H.C

# Supplement for "Elastocaloric signature of the excitonic instability in Ta$_2$NiSe$_5$".

## A. Elastocaloric measurements along the a-axis (A$_{1g}$)

Figure S1 depicts the elastocaloric effect for the system but with uniaxial stress applied along the a-axis, i.e the Ta chain direction. Strain induced in this orientation does not change any point group symmetries and so has purely A$_{1g}$ character, but it can affect bandgaps/overlaps in the system regardless of excitonic contributions. It is clear this measurement is still sensitive to the phase transition, with the bump at 333K (different from the sample presented in the main text likely due to non-ideal thermalization) present for all offset DC strains likely arising from the relationship:

$$\frac{dT}{d\varepsilon_{A_{1g}}} = \frac{dT_S}{d\varepsilon_{A_{1g}}} \frac{C_p^{crit}}{C_p^{tot}}$$

The jump corresponding to the heat capacity feature is nonconstant with A$_{1g}$ offset strain, indicating either non-linearities in $\frac{dT_S}{d\varepsilon_{A_{1g}}}$ or $C_p^{crit}\left(\varepsilon_{A_{1g}}\right)$.

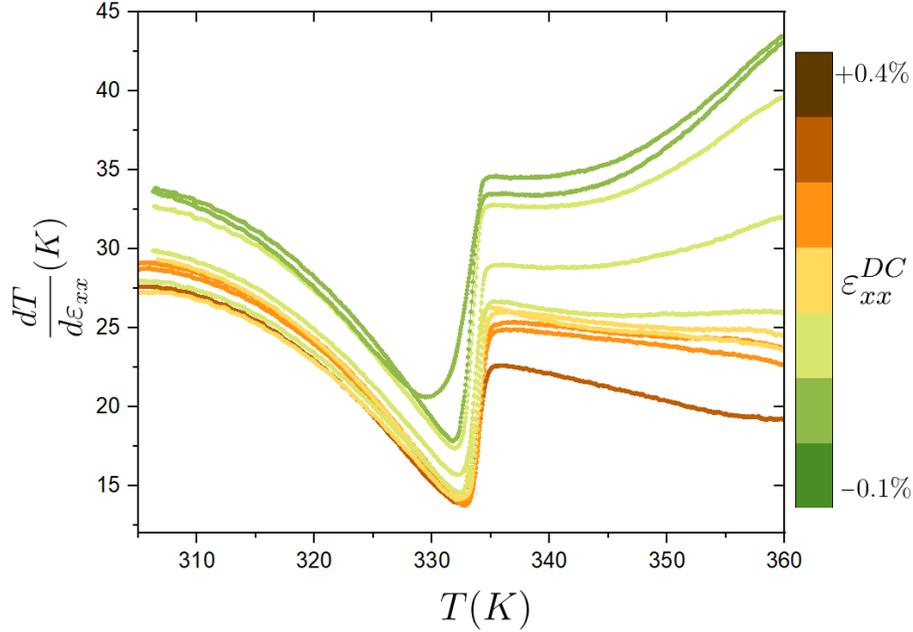

**Figure S1: Elastocaloric effect at different offset strains, probing the A$_{1g}$ channel.** Plotted is the magnitude of the temperature oscillations induced from a small AC strain (≈ 0.005%) the sample experienced, which manifests from the elastocaloric effect. Uniaxial AC stress was applied along a-axis while sweeping temperature from 300K to 360K, for different DC stresses which induced DC strains estimated to range from 0.4% (in brown) to -0.1% (in green).

Interestingly the ECE for induced A$_{1g}$ strain is non-constant and large both below and above the phase transition temperature. This indicates there are sizable contributions to the free energy for the system containing $\varepsilon_{A_{1g}}$ terms. Potentially the direct bandgap is also tuned by A$_{1g}$ strain. Importantly though, the switching of sign and divergence seen in the B$_{2g}$ channel was not observed for this channel, indicating the uniqueness of B$_{2g}$ strain in the excitonic/entropy landscape of Ta$_2$NiSe$_5$.

## B. Considering the free energy: optical phonons, excitonic fluctuations, and strain

It is important to note that the free energy discussed in Eq. 1 in the main text considers only one generic non-acoustic order parameter, while multiple previous experiments have pointed to the fact that two separate non-acoustic degrees of freedom, optical phonons and excitonic fluctuations, must be considered separately. For Ta$_2$NiSe$_5$ this would involve a **q=0** B$_{2g}$ optical phonon coupling to the shear along the ac plane. The amplitude of the displacement mode **Q** therefore bilinearly couples to the lattice strain $\varepsilon_{B_{2g}}$. Similarly an order parameter representing excitonic condensation also bilinearly couples to both strain and the optical phonon amplitude mode. We continue by writing a free energy in which both of these order parameters are allowed to have the generic temperature dependence $a(T - T^*)$ for driving the phase transition (although this need not be the case, one of them may be temperature independent), but emphasize that we do not include this for the elastic energy term as our elastocaloric measurements have made it clear that the non-acoustic degrees of freedom drive the phase transition:

$$F_{\psi_e,Q,\varepsilon} = \frac{a_e}{2}(T - T_e)\psi_e^2 + \frac{a_O}{2}(T - T_O)Q^2 - \lambda_{eO}\psi_e Q - \lambda_{eS}\psi_e \varepsilon_{B_{2g}} - \lambda_{OS}Q\varepsilon_{B_{2g}} + \frac{C_{55}^o}{2}\varepsilon_{B_{2g}}^2 + O(\psi_e^4, Q^4)$$

T$_e$ would be the bare transition temperature if only excitons were involved, similarly T$_O$ for optical phonons, and $\lambda_{ij}$ are the corresponding coupling constants. $C_{55}^o$ is the bare elastic constant in the B$_{2g}$ symmetry.

We note that the quadratic coefficient of the displacement **Q** is proportional to the square of the optical phonon frequency: $a_O(T - T_O) = m\omega^2$, and so the characteristic measure to probe this term is a softening of the phonon frequency towards 0 at temperatures above the phase transition, which can be detected in measurements like Raman spectroscopy.

Here we remark on three relatively robust experimental results for this compound:

1. Several sets of Raman measurements have not seen an appreciable softening of the B$_{2g}$ optical phonon mode towards T$_S$. This implies A(T) is not very singular, implying T$_O$ is not particularly large compared to T$_S$.

2. Recent time-resolved ARPES measurements have nonetheless shown that phonons do play a crucial role in the phase transition, as the relaxation time scales for the gap arising from the order parameter are shown to be from structural degrees of freedom rather than electronic/excitonic.

3. These elastocaloric measurements establish that it is highly likely the non-acoustic degrees of freedom (coupled optical phonon/electronic-excitonic continuum) drive the phase transition. This also indicates that both couplings to the lattice: $\lambda_{eS}$ and $\lambda_{OS}$, are small in magnitude.

To more explicitly demonstrate this let's consider the elastocaloric results by first minimizing the free energy with respect to $\psi_e$ and substituting it in terms of Q and $\varepsilon$.

$$F_{Q,\varepsilon} = \left(\frac{a_O}{2}(T - T_O) - \frac{\lambda_{eO}^2}{2a_e(T - T_e)}\right)Q^2 - \left(\frac{\lambda_{eS}\lambda_{eO}}{2a_o(T - T_e)} + \lambda_{OS}\right)Q\varepsilon_{B_{2g}} + \frac{C_{55}^o}{2}\varepsilon_{B_{2g}}^2$$

We note had we instead minimized for Q we would just replace the subscript e for O for the coefficients in the free energy. For simplicity we define these temperature dependent coefficients as:

$B(T) = \left(\frac{\lambda_{eS}\lambda_{eO}}{2a_e(T-T_e)} + \lambda_{OS}\right)$ and $A(T) = \left(a_O(T - T_O) - \frac{\lambda_{eO}^2}{a_e(T-T_e)}\right)$.

If we approximate A(T) as $A = a_o + \lambda^2/a_e(T - T_e)$ (matching the lack of softening observed) we can then solve for the term $\frac{dS}{d\varepsilon}$ to obtain:

$$\frac{dS}{d\varepsilon_{B_{2g}}} = \frac{d}{d\varepsilon_{B_{2g}}}\frac{dF}{dT} - \varepsilon_{B_{2g}}\frac{d}{dT}\frac{B(T)}{A(T)} = \varepsilon_{B_{2g}}\frac{d}{dT}\left(\frac{\lambda_{eS}\lambda_{eO} + \lambda_{OS}(2a_e(T - T_e))}{a_o a_e(T - T_e) - \lambda_{eO}^2}\right)$$

Here if $T - T_e \approx T$ we can safely ignore the temperature dependence of the second term in the temperature range we measured (from 315K to 370K) of the numerator and we are left with:

$$\frac{dS}{d\varepsilon_{B_{2g}}} = -\varepsilon_{B_{2g}} \frac{d}{dT}\left(\frac{\lambda_{eS}\lambda_{eO} + C\lambda_{OS}}{a_o a_e (T - T^*)}\right)$$

Where $T^* = T_e + \lambda_{eO}^2/a_o a_e$. Thus the elastocaloric measurements will measure this renormalized quantity T*. Now the phase transition temperature $T_S$ occurs when the renormalized quadratic coefficient of the acoustic strain goes to 0 (as there is a spontaneous strain observed below $T_S$. This occurs when:

$$\frac{C_{55}^o}{2} = \frac{B(T)}{2A(T)} = \frac{\lambda_{eS}\lambda_{eO} + \lambda_{OS}(2a_e(T - T_e))}{2a_o a_e (T - T_e) - 2\lambda_{eO}^2}$$

This can be solved for $T_S$:

$$T_S = T_e + (T^* - T_e)\left(\frac{C_{55}^o}{C_{55}^o - 2\lambda_{OS}/a_o}\right)\frac{\lambda_{eO}\lambda_{eS}}{a_e a_o (C_{55}^o - 2\lambda_{OS}/a_o)}$$

Our elastocaloric measurements set strict limits on these terms: because $(T_S - T^*)/T_S$ was measured to be much smaller than $T^*/T_S$, both $\lambda_{OS}$ and $\lambda_{eS}$ must be very small.

Thus the combination of previous Raman measurements, time-resolved ARPES measurements, and these elastocaloric measurements unambiguously signify that the electron/exciton-optical phonon coupling plays a crucial role in driving this phase transition, and that all other couplings and bare transition temperatures must be significantly energetically lower than $T_S$.

### C. Calculated Poisson Ratio

For an experimental geometry in which uniaxial stress is applied along (1 0 0), the corresponding strains are induced via the compliance tensor terms:

$$\varepsilon_i = S_{i1}\sigma_1$$

Where the Voigt notation is being used, which for an orthorhombic system is 1=xx, 2=yy, 3=zz, 4=yz, 5=xz, 6=xy, and x is along the a-axis, y is along the b-axis, and z is along the c-axis. To determine the in-plane Poisson ratio, which will help indicate the relative amount of $B_{2g}$ vs. $A_{1g}$ strain induced at any given temperature, the relevant equation is:

$$\gamma = \frac{S_{31}}{S_{11}} = \frac{C_{13}C_{22} - C_{23}C_{12}}{C_{33}C_{22} - C_{23}^2}$$

However for the measurements presented in this work, the uniaxial stress was applied along 45 degrees from the a-axis in the a-c plane. Hence the previous equation should be rotated where $x' = \sqrt{2}/2\hat{x} + \sqrt{2}/2\hat{z}$ and $z' = -\sqrt{2}/2\hat{x} + \sqrt{2}/2\hat{z}$. Rotating the compliance tensor terms gives:

$$\gamma' = \frac{C_{22}\left(C_{A_{1g}} - C_{55}\right) - C_{12}'^2}{C_{22}\left(C_{A_{1g}} + C_{55}\right) - C_{12}'^2}$$

Where $C_{A_{1g}} = (C_{11} + C_{33} + 2C_{13})/4$ and $C_{12}' = (C_{12} + C_{23})/2$. For a material undergoing a structural phase transition in the $B_{2g}$ channel, $C_{55}$ is expected to soften to 0 at $T_S$. In fact recent ultrasound experiments of $(Ta_{0.952} V_{0.048})_2 NiSe_5$ show that $C_{55}$ softens to nearly 90% of its value at $T_S$. This softening will create a temperature-dependent Poisson ratio, which will in turn produce a temperature dependent $B_{2g}$ strain transmission ratio relative to the measured $\varepsilon_{xx}$. To attempt to account for this the relevant elastic constant values were taken from Ref S1 except for $C_{13}$, $C_{12}$, and $C_{23}$ which were given values of $5\times10^{10}$ J/m^3, and $C_{55}$, which is explained below.

The elastocaloric effect, via effective Maxwell relations, can be related to the elastic constants of the material for the relevant symmetry channel(s) i:

$$\left(\frac{dT}{d\varepsilon_i}\right)_S = \varepsilon_i \left(\frac{T}{C_\varepsilon}\right)\frac{dC_i}{dT}$$

as discussed in Ref S2. Thus if a certain T* is obtained for the Curie-Weiss form of the elastocaloric effect, the same temperature should be the relevant T* in the form of the softening elastic constant. Thus the form:

$$C_{55}(T) = C_o - C_o\left(\frac{T - (T_S - T^*)}{T - T^*}\right)$$

was plugged into the Poisson ratio with various T*, and the best matches between the proposed T* and the one produced by the fit occurred when T*=298K. The data for the extracted temperature derivative of the $\varepsilon_{xx}$ susceptibility, which does not take into account softening of the $B_{2g}$ channel, is shown in Figure S3 with a similar but slightly higher Weiss temperature (307K) as expected.

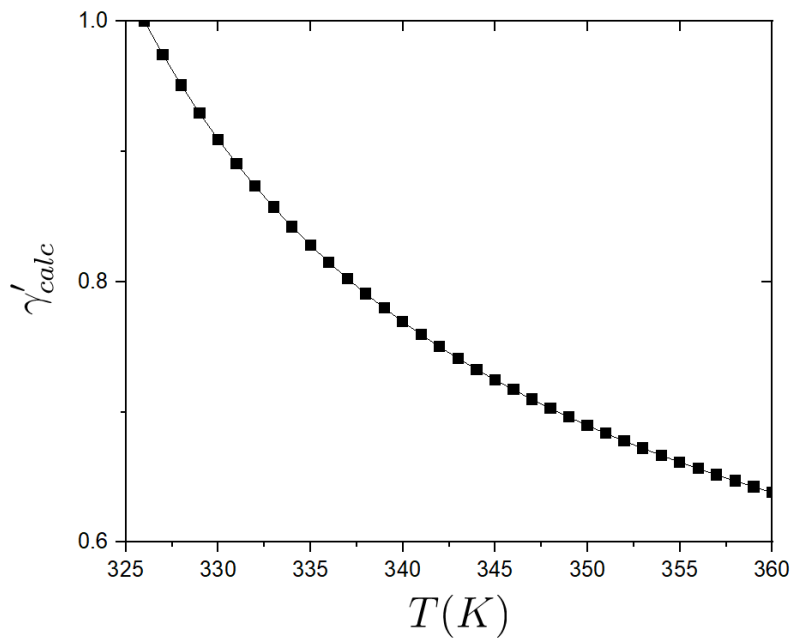

**Figure S2:** Calculated Poisson ratio which was used to determine strain transmission ratios.

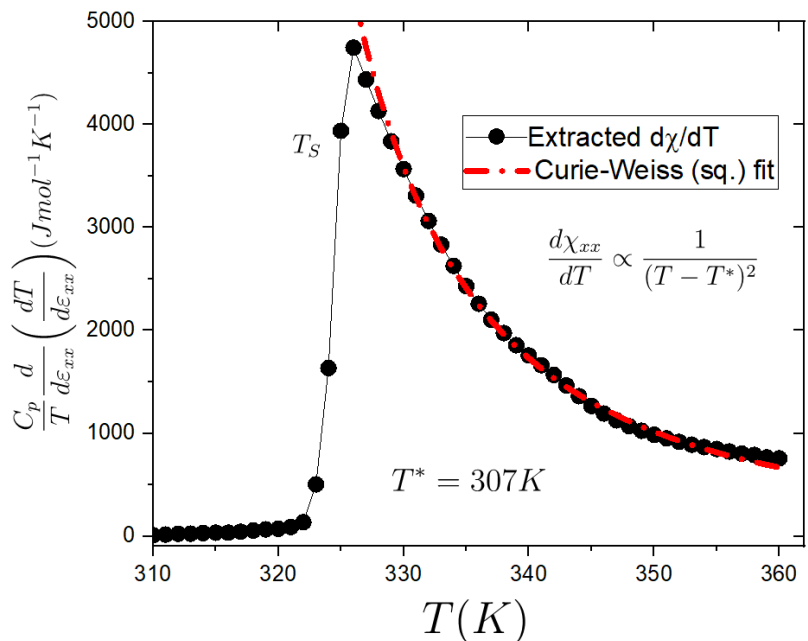

**Figure S3: DC strain derivative of the normalized elastocaloric effect**. This data only looks at the derivatives with respect to $\varepsilon_{xx}$ and hence overestimates T* because the softening of the lattice implies the relative fraction of $B_{2g}$ strain the sample experiences increases as the system cools towards $T_S$.

**D. Transport**

A resistivity measurement of a free-standing Ta$_2$NiSe$_5$ single crystal grown was performed with the transport direction along the a-axis to confirm the crystal quality. The results of this are displayed in Figure S4 as well as the extracted transport gap $\Delta = k_B T^2 \frac{d \ln(\rho)}{dT}$, assuming the resistivity follows the relation: $\rho = \rho_o e^{\Delta/k_B T}$. This is consistent with previously measured transport on this compound.[S3]

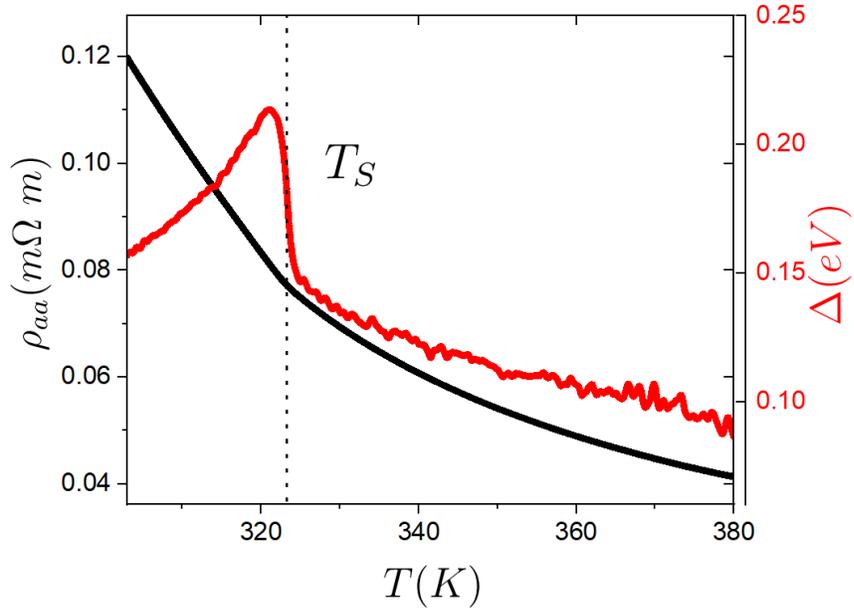

**Figure S4. Freestanding resistivity measurement along the a-axis, and the corresponding extracted gap.**

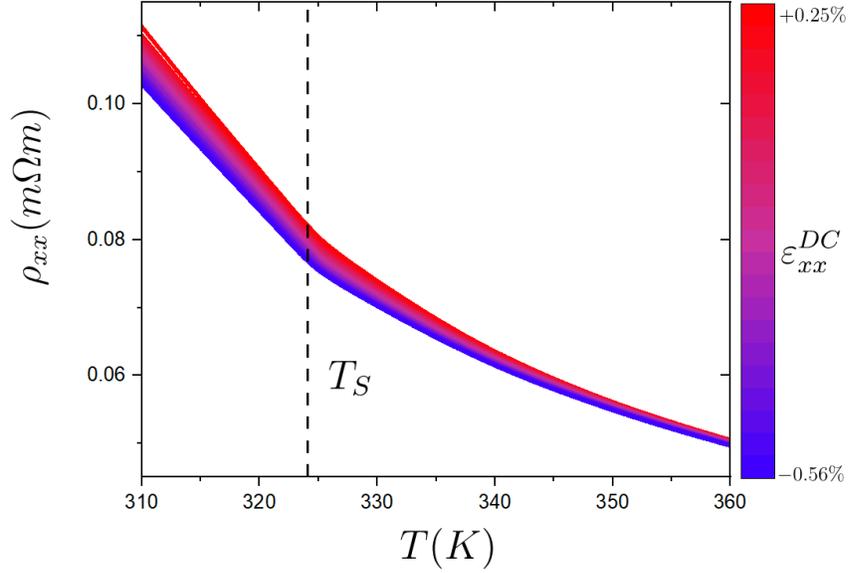

**Figure S5. Resistivity data at different offset strains.** Plotted is the resistivity measured sweeping temperature from 310K to 360K, for different DC stresses (applied and measured at 310K) which induced strains ε$_{xx}$ estimated to range from 0.25% (in red) to -0.56% (in blue).

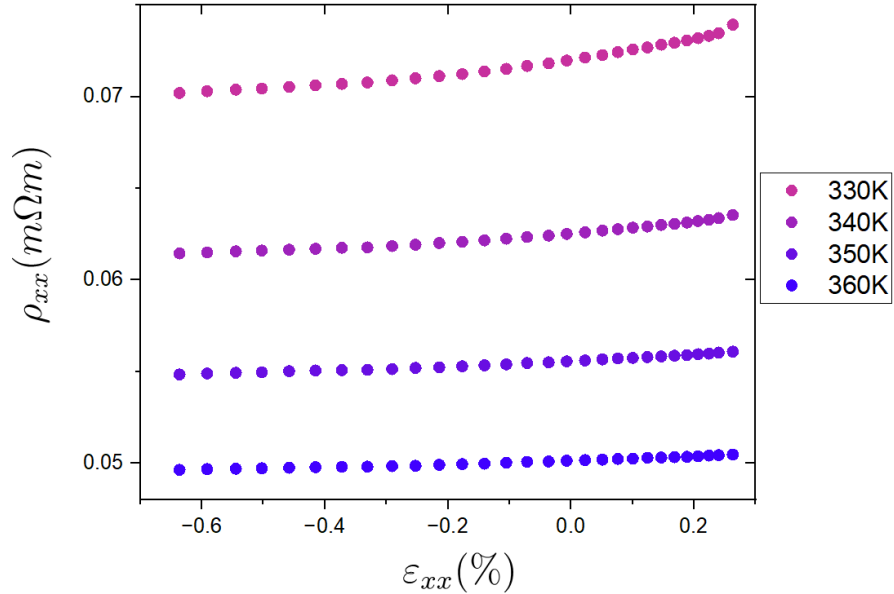

**Figure S6. Resistivity binned at various temperatures, plotted against DC strain.** Note it is non-linear at all temperatures, with increasing non-linearity as the phase transition is approached.

Resistivity measurements were also performed simultaneously with the elastocaloric measurements presented in the main text, with the current and voltage directions along the same axis the uniaxial stress was applied (45 degrees from the a-axis). The results are displayed in Figure **S5**. This data was binned at select temperatures, and is plotted against strain in Figure **S6**.

For the geometry of the measurement only ρ$_{xx}$ was measured, which along 45 degrees from the a-axis gives the relevant resistivity tensor elements:

$$\rho_{xx} = \frac{(\rho_{aa} + \rho_{cc})}{2} + \rho_{ac}.$$

Although resistivity is not a thermodynamic quantity, its dependence on strain can also shed light on the nature of the order parameter and its high-temperature fluctuations. As has been previously shown[S3], resistivity is sensitive to the gap that results from the structural phase transition. Usually for semiconductors, the size of the bandgap is linearly modified by strain, and so it is common to have a large first order derivative of resistivity with respect to strain, known as the linear elastoresistivity coefficient. Much less common is a significant *non-linear* elastoresistivity, indicating the transport gap is being tuned quadratically with $B_{2g}$ strain at temperatures above $T_S$.